\documentclass{article}
\usepackage{spconf, amsmath, graphicx}
\usepackage{multirow}
\usepackage{algorithm,algpseudocode}

\usepackage{url}
\usepackage{hyperref}

\usepackage[noadjust]{cite} 

\usepackage[utf8]{inputenc}

\title{Latent Filling: Latent Space Data Augmentation \\ for Zero-shot Speech Synthesis}
\name{Jae-Sung Bae$^{\dagger}$, Joun Yeop Lee$^{\dagger}$, Ji-Hyun Lee$^{\dagger}$, Seongkyu Mun$^{\dagger}$, Taehwa Kang$^{\dagger}$, Hoon-Young Cho$^{\dagger}$,}
\sname{Chanwoo Kim$^{\ddagger}$\sthanks{Work performed while at Samsung Research.}}
\address{$^{\dagger}$Samsung Research, Seoul, Republic of Korea \\ $^{\ddagger}$Korea University, Seoul, Republic of Korea}

\begin{document}
\ninept
\maketitle
\begin{abstract}
Previous works in zero-shot text-to-speech (ZS-TTS) have attempted to enhance its systems by enlarging the training data through crowd-sourcing or augmenting existing speech data. However, the use of low-quality data has led to a decline in the overall system performance. To avoid such degradation, instead of directly augmenting the input data, we propose a latent filling (LF) method that adopts simple but effective latent space data augmentation in the speaker embedding space of the ZS-TTS system. By incorporating a consistency loss, LF can be seamlessly integrated into existing ZS-TTS systems without the need for additional training stages. Experimental results show that LF significantly improves speaker similarity while preserving speech quality.
\end{abstract}
\section{Introduction}
\label{sec:intro}
With the advancements in neural text-to-speech (TTS) systems \cite{tacotron2, fastspeech2, lpctron}, there has been a remarkable improvement in the naturalness of synthesized speech. There is also a growing demand for personalized TTS systems. Building such systems traditionally involves adapting pre-trained TTS systems using a limited number of utterances from the target voice \cite{adapt1, adapt2}. However, the process of collecting and fine-tuning with personal data remains problematic due to privacy concerns and the challenges associated with gathering high-quality data.

To address this challenge, zero-shot TTS (ZS-TTS) systems have gained significant attention \cite{yourtts, transfer, sc-glowtts, adaspeech4, lee2022pvae, snac, lee23f_interspeech} as they aim to replicate a target speaker's voice using just a single reference utterance without the need for additional fine-tuning. A common architecture employed in ZS-TTS systems incorporates an external speaker encoder \cite{sc-glowtts, yourtts}, which is pre-trained on a speaker verification task to extract speaker embeddings from reference speeches. To effectively interpret the latent space of speaker embeddings and generate speech from an unseen speaker embedding during inference, ZS-TTS systems require substantial training data encompassing a diverse set of speakers. However, acquiring high-quality speech-text paired data for training is a costly and time-consuming endeavor.

To overcome the data scarcity challenge, recent TTS systems have utilized crowd-sourced speech data \cite{valle, spear-tts} or employed data augmentation techniques such as pitch shifting \cite{pitch-shift} and synthesizing new speech using voice conversion or TTS systems \cite{pitch-shift, vc-dataaug, tts-by-tts-2}. Nevertheless, these data sources often contain speech with ambiguous pronunciation, background noise, channel artifacts, and artificial distortions, which result in degradation of the overall performance of the TTS systems.

In this paper, we propose a novel approach called \textit{latent filling} (LF) to address these challenges. LF aims to \textit{fill} the unexplored regions of the latent space of speaker embeddings through latent space data augmentation \cite{latent_aug1, mixup, modals, latent_aug2}. Unlike data augmentation techniques applied directly to input data, latent space data augmentation is a straightforward yet effective method that augments the latent vectors of neural network systems, enhancing their robustness and generalization. Although widely used in classification tasks \cite{mixup, modals, latent_aug2}, latent space data augmentation has not been extensively explored in generation tasks due to the inherent difficulty of obtaining corresponding target data for augmented latent vectors.

To tackle this challenge, we introduce a latent filling consistency loss (LFCL). LFCL enforces the generated acoustic features derived from augmented speaker embeddings to retain the same speaker embedding representation. This allows us to train the entire TTS system without requiring the corresponding target speech sample for the augmented speaker embedding. Our approach involves training the ZS-TTS system in two modes. First, when LF is adopted, the entire network is exclusively trained with LFCL. Conversely, when LF is not applied, we employ a reconstruction loss and speaker consistency loss \cite{yourtts}. By randomly selecting these modes during training iterations, we successfully integrate LF into the training of the ZS-TTS system without the need for additional training stages.

With minimal modifications to the existing code, LF can be easily applied to existing ZS-TTS systems. Furthermore, our experiments demonstrate that the incorporation of LF improves the speaker similarity performance of the ZS-TTS system without any degradation in intelligibility and naturalness.

\begin{figure}[t]
    \centering
    \centerline{\includegraphics[width=0.8\linewidth]{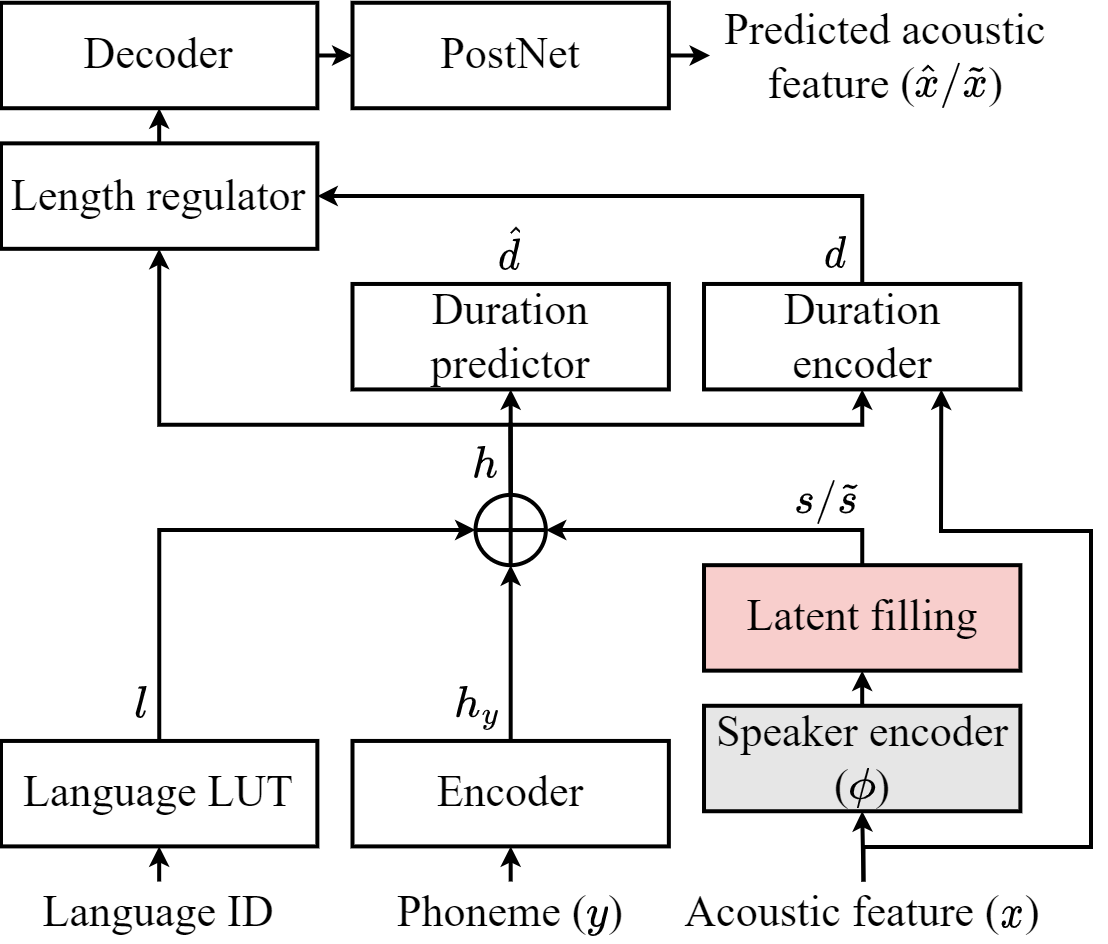}}
    \caption{Architecture of the baseline ZS-TTS system with our proposed latent filling method.}
\label{fig:model}
\end{figure}

\section{Proposed Method}
\label{sec:model}

\subsection{Baseline ZS-TTS system}
Our baseline ZS-TTS system shares a similar architecture with \cite{lee23f_interspeech}. It is designed for low-resource and cross-lingual speech generation. The overall architecture is illustrated in Fig. \ref{fig:model}. The input phoneme $y$ is encoded into $h_y$ by an encoder. Language information is transformed into a language embedding $l$ via a look-up table (LUT). A pre-trained speaker encoder $\phi$ extracts the speaker embedding $s$ from the input acoustic feature $x$. Notably, the speaker encoder remains frozen during the training phase. Subsequently, $h_y$, $l$, and $s$ are concatenated to form the final hidden representation $h$. To extract the corresponding duration $d$ from $x$ and $h$, the duration encoder and alignment method from \cite{align} are employed. The duration predictor predicts $\hat{d}$, which is utilized as duration information during inference. Finally, $h$ is upsampled using the duration information and passed through the decoder and PostNet to generate the final acoustic feature $\hat{x}$. 

During the training phase, the system is trained with reconstruction loss, denoted as $L_{Rec}$, for the duration and acoustic features. L1 and L2 losses are used for the duration and acoustic feature reconstruction loss, respectively. Additionally, as in \cite{scl2} and \cite{yourtts}, we employ the speaker consistency loss (SCL) to enhance the speaker similarity of the ZS-TTS system. It encourages the speaker embedding of the generated acoustic feature to be close to the input speaker embedding. We calculate the cosine similarity to measure the similarity between two speaker embeddings. With batch size $N$, the SCL is computed as follows:
\begin{equation}
    L_{SCL} = -\frac{1}{N} \sum^N_i{\textit{cos\_sim} (s_i, \phi(\hat{x_i}))}
\end{equation}
\label{loss:scl}

\begin{figure}[t]
\centering
{
    \centering
    \begin{minipage}[t]{.49\linewidth}
      \centering
      \centerline{\includegraphics[width=0.8\linewidth]{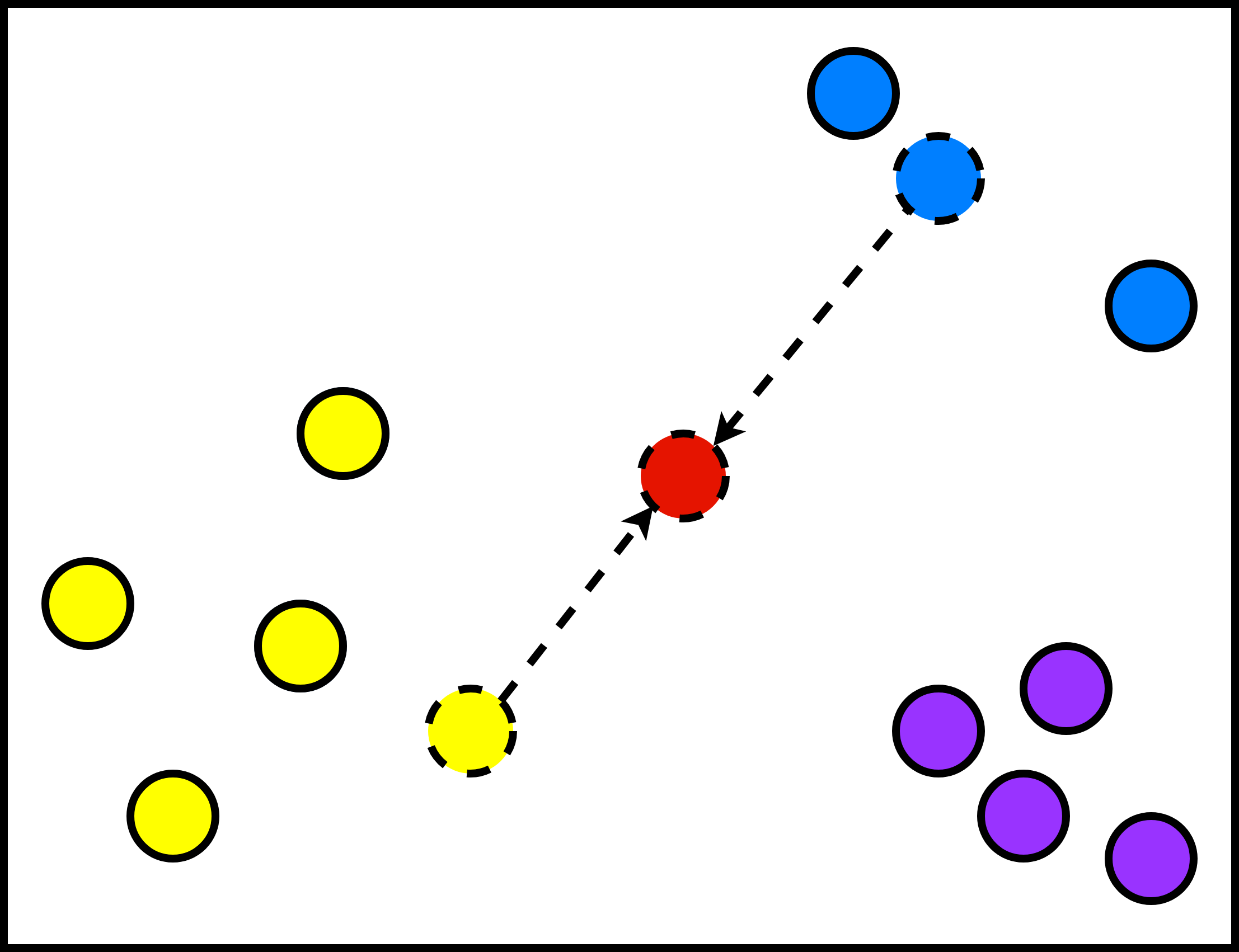}}
      \centerline{(a)}
    \end{minipage}
    \centering
    \begin{minipage}[t]{.49\linewidth}
      \centering
      \centerline{\includegraphics[width=0.8\linewidth]{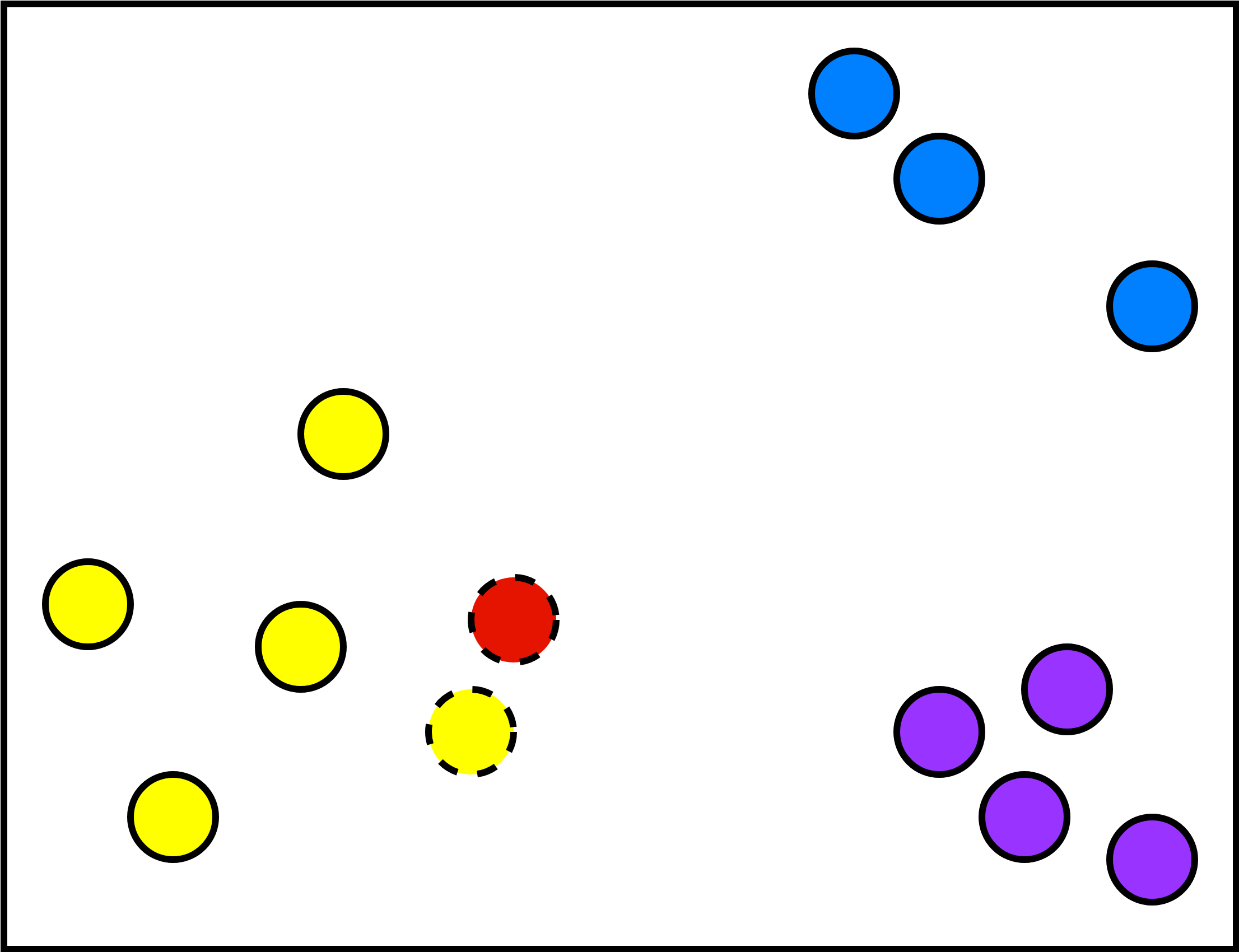}}
      \centerline{(b)}
    \end{minipage}
    \caption{Illustration of LF of (a) interpolation and (b) noise adding. The red circle indicates an augmented speaker embedding, while the circles in various colors represent speaker embeddings of different speakers.}
    \label{fig:lf}
}
\end{figure}

\begin{figure}[t]
\vspace{-0.3cm}
\newlength{\continueindent}
\setlength{\continueindent}{2em}
\makeatletter
\newcommand*{\ALG@customparshape}{\parshape 2 \leftmargin \linewidth \dimexpr\ALG@tlm+\continueindent\relax \dimexpr\linewidth+\leftmargin-\ALG@tlm-\continueindent\relax}
\makeatother

\renewcommand{\algorithmicensure}{\textbf{Begin}} 
\newcommand{\Beegin}{\textbf{Beegin}} 
\begin{algorithm}[H] 
   \caption{Latent filling algorithm}\label{algo_proposed} 
  \begin{algorithmic}[1]
  \Require Speaker embedding $s_i$ and $s_j$. 
  \Require Noise adding probability $\epsilon \in [0, 1]$.
  \Require Beta distribution shape parameter $\beta \in(0,\infty)$.
  \Require Standarad deviation of Gaussian noise $\sigma \in (0,\infty)$.
    \State $u_1, u_2 \sim \text{U}(0,1), \lambda \sim \text{Beta}(\beta, \beta), G \sim \mathcal{N}(0, \sigma^2)$
    \If{$u_1 > \epsilon$} 
    \State $\tilde{s_i} \gets \lambda s_i + (1 - \lambda)s_j$ \Comment{Perform interpolation}
    \If{$u_2 < \epsilon$} 
    \State $\tilde{s_i} \gets \tilde{s_i} + G$  \Comment{Add Gaussian noise}
    \EndIf
    \Else{}
    \State $\tilde{s_i} \gets s_i + G $ \Comment{Add Gaussian noise} 
    \EndIf
    \State \Return $\tilde{s_i}$
  \end{algorithmic}
\end{algorithm}
\end{figure}

\subsection{Latent Filling}

Through the latent filling (LF) method, our aim is to \textit{fill} the latent space of the speaker embeddings that the training dataset cannot adequately express. We employ two intuitive latent space augmentation techniques for the LF: interpolation \cite{latent_aug1, mixup} and noise addition \cite{latent_aug1}. Illustrations of these methods are provided in Fig. \ref{fig:lf}. The interpolation method creates a completely new speaker embedding in the speaker embedding space by using two different speaker embeddings, while the adding noise method generates a new speaker embedding that is relatively close to the existing one. In the previous work \cite{latent_aug1}, the authors also adopted the extrapolation for latent space data augmentation. However, in our preliminary experiments, we observed that extrapolation is not particularly meaningful in our case.

The complete LF process is detailed in Algorithm \ref{algo_proposed}. First, for the speaker embedding $s_i$, we randomly select another speaker embedding $s_j$ from the training dataset for the interpolation. Our preliminary study indicates that ensuring $s_j$ has the same language information as $s_i$ is crucial for achieving stable performance. We perform interpolation between $s_i$ and $s_j$ with a probability of $1-\epsilon$, using the formula $\lambda s_i + (1 - \lambda)s_j$. Here, $\epsilon$ represents the probability of adding noise, while $\lambda$ denotes the interpolation rate. Similar to \cite{mixup}, we sample $\lambda$ from the beta distribution Beta$(\beta, \beta)$, where $\beta$ is the beta distribution shape parameter. When interpolation is conducted, Gaussian noise $G\sim\mathcal{N}(0, \sigma^2)$ with a standard deviation $\sigma$ is added with a probability of $\epsilon$. Conversely, when the interpolation is not performed, the Gaussian noise is always added.

\subsubsection{Latent filling consistency loss} 
Adopting latent space data augmentation for generation tasks has been challenging due to the inherent difficulty of obtaining target data corresponding to augmented latent vectors. Similarly, when the LF method is adopted for the speaker embedding, it is impossible to calculate the $L_{Rec}$ because there is no corresponding ground-truth speech containing the speaker information for the augmented speaker embedding $\tilde{s}$. 

To address this challenge, we propose a latent filling consistency loss (LFCL), a modified version of SCL tailored for augmented speaker embeddings. The LFCL can be computed as follows:
\begin{equation}
    L_{LFCL} = -\frac{1}{N} \sum^N_i{\textit{cos\_sim} (\tilde{s_i}, \phi(\tilde{x_i}))}
\end{equation}
\label{loss:lfcl}where $\tilde{x}$ is a generated acoustic feature corresponding to $\tilde{s}$. LFCL measures the closeness between the speaker embedding of $\tilde{x}$ and $\tilde{s}$, and encourages $\tilde{x}$ to have the same speaker characteristics as the input augmented speaker embedding $\tilde{s}$. By using the LFCL, we can successfully update the ZS-TTS system with the augmented speaker embedding without the need for $L_{REC}$. This allows the ZS-TTS system to be trained with speaker embeddings not contained in the training dataset. The training process incorporating LFCL is detailed in the following section.

\subsubsection{Training procedure}
The training procedure for the TTS system using the LF and LFCL is as follows: First, for each training iteration, we make a random decision on whether to perform the LF, with a probability parameter $\tau$ ranging from 0 to 1. When the LF is performed, the entire TTS system is only updated with $L_{LFCL}$. Conversely, when LF is not applied, the TTS system is updated using  $L_{Rec}$ and $L_{SCL}$. By randomly incorporating LF during training, we seamlessly integrate it into existing TTS systems without requiring additional training stages or degrading performance.

According to our preliminary experiments, the parameter $\tau$ should be set carefully. When $\tau$ was set too high, the TTS system generated speech that lacked coherence with the input text. This occurred because $L_{Rec}$ is not utilized when LF is active, leading the system to prioritize expressing speaker characteristics of the input speaker embedding while disregarding content information. Conversely, when $\tau$ was set too low, the benefits of LF were not effectively realized. Based on heuristic analysis, we set $\tau$ to a value of 0.25.

\section{Experiments}
\label{sec:exp}

\subsection{Dataset}
\label{ssec:dataset}
In our experiments, we utilized a diverse set of datasets for both English and Korean languages. For English, we used VCTK \cite{vctk}, LibriTTS \cite{libritts} (train-clean-100 and 360 subsets), and LJSpeech \cite{lj} dataset, totaling approximately 259 hours of speech data and involving 1,245 distinct speakers. We excluded 11 speakers from the VCTK dataset and used them as English test speakers. For Korean language, we leveraged the multi-speaker\footnote{https://aihub.or.kr/aihubdata/data/view.do?dataSetSn=542} and emotion\footnote{https://aihub.or.kr/aihubdata/data/view.do?dataSetSn=466} datasets from the AIHub. They were recorded by 3,086 non-professional speakers and 44 professional voice actors, and the lengths of each dataset were approximately 7,414 hours and 264 hours, respectively. We randomly selected 29 speakers from the AIHub multi-speaker dataset for use as Korean test speakers. We resampled all utterances to a 24KHz sampling rate. Our test scripts were derived from the utterances of English test speakers, with a focus on those between 4 and 10 seconds in duration. 

We used 22-dimensional acoustic features consisting of 20 Bark cepstral coefficients, pitch period, and pitch correlation which were the same as in \cite{lee23f_interspeech}. To convert the acoustic feature to a waveform, Bunched LPCNet2 \cite{lpcnet} was used.

\subsection{Experimental setup}
The baseline ZS-TTS system had the same architecture as our previous work \cite{lee23f_interspeech}, except for the language LUT and the decoder architecture. The dimension of $l$ was 4 and the Uni-LSTM was used for the decoder instead of the Bi-LSTM to enable streaming. For the LF, we set $\epsilon$ to 0.5, $\beta$ to 0.5, and $\sigma$ to 0.0001. For the speaker encoder, ECAPA-TDNN \cite{ecapa}, the state-of-the-art (SOTA) speaker verification model, was used. To compute the SCL and LFCL, we modified the input acoustic feature of the ECAPA-TDNN to ours. It was pre-trained with the same training dataset. Our system was trained with a batch size $N$ of 64. We adopt the language-balanced sampling following \cite{xlsr} and \cite{voicebox} with the upsampling factor of 0.25 to ensure a language-balanced batch. We used the Adam \cite{adam} optimizer with betas 0.9 and 0.999, weight decay of $10^{-6}$, and an initial learning rate of 0.0006 with a half-learning rate schedule for every 100K iterations. The system was trained for 1 million iterations. To generate speech in a zero-shot scenario, we randomly selected one utterance with a duration longer than 4 seconds from each test speaker as a reference speech.

\subsection{Comparison Systems}
For the comparison, we used the following systems. \textbf{GT} and \textbf{GT-re} were the ground-truth speech samples and re-synthesized version of GT through the vocoder, respectively. \textbf{SC-Glow TTS} \cite{sc-glowtts} and \textbf{YourTTS} \cite{yourtts} were the open-sourced ZS-TTS systems. Furthermore, we compared the proposed LF method (\textbf{Baseline+LF}) with data augmentation methods. The \textbf{Baseline+CS} system utilized LibriLight \cite{librilight} dataset, which is a large amount of crowd-sourced data that consists of a total of 60K hours of unlabelled speech with approximately 7,000 speakers. To generate transcripts, we used a wav2vec \cite{xlsr}-based phoneme recognition model. We also built \textbf{Baseline+PS} system which utilized pitch shifts to increase the amount and diversity of the training data. We applied pitch shifts to each speech in the training dataset and doubled the amount of training data. PRAAT toolkit \cite{praat} within the semitone shift range [-4, 4] was used for pitch shift.

\subsection{Evaluation metrics}
For the objective test, we evaluate the speaker similarity and the intelligibility of each system with averaged speaker embedding cosine similarity (SECS) and word error rate (WER), respectively. The speaker encoder of the Resemblyzer\footnote{https://github.com/resemble-ai/Resemblyzer} package was used to measure the SECS. The SECS ranges from -1 to 1, and a higher score implies better speaker similarity. To compute the WER, Whisper \cite{whisper}, an open-source automatic speech recognition (ASR) model, was used. 

For the subjective evaluation, we conducted two mean opinion score (MOS) tests; a MOS test on overall naturalness of speech, and a speaker similarity MOS (SMOS) test that focused on evaluating the speaker similarity between generated speech and reference speech. Testers of both tests were requested to evaluate speech in the range from 1 to 5 with an interval of 0.5, where 5 is the best. For the subjective tests, 90 testers participated via Amazon MTurk. 

\begin{table*}[t]
    \small
    \centering
    \begin{tabular}{l|r|r|r|r|r|r|r|r|r}
    \hline
    \multirow{2}{*}{\textbf{System}} & \multicolumn{1}{c|}{\multirow{2}{*}{\textbf{NP}}} & 
    \multicolumn{4}{c|}{\textbf{Intra-lingual (En $\rightarrow$ En)}} & \multicolumn{4}{c}{\textbf{Cross-lingual (Ko $\rightarrow$ En)}}\\
    \cline{3-10}
    & & \multicolumn{1}{c|}{\textbf{SECS}}  & \multicolumn{1}{c|}{\textbf{WER ($\%$)}} & \multicolumn{1}{c|}{\textbf{MOS}}  & 
    \multicolumn{1}{c|}{\textbf{SMOS}} & 
    \multicolumn{1}{c|}{\textbf{SECS}}  & \multicolumn{1}{c|}{\textbf{WER ($\%$)}} & \multicolumn{1}{c|}{\textbf{MOS}}  & 
    \multicolumn{1}{c}{\textbf{SMOS}} \\
    \hline
     GT                   & - & 0.882 & 1.95 & 4.12 $\pm$ 0.09 & 3.89 $\pm$ 0.11  & (0.869) & - & - & (3.66 $\pm$ 0.10) \\
     GT-re                & -& 0.855 & 3.32 & 3.77 $\pm$ 0.14 & 3.37 $\pm$ 0.15 & (0.836) & - & - & (3.29 $\pm$ 0.14) \\
    \hline
    SC-Glow TTS           &30.1M& 0.599 &28.56 & 2.02 $\pm$ 0.14 & 2.08 $\pm$ 0.17 & 0.667 & 29.00 & 2.06 $\pm$ 0.14 & 2.12 $\pm$ 0.14\\
    YourTTS               &40.1M& 0.810 & 4.78 & 3.16 $\pm$ 0.16 & 2.79 $\pm$ 0.16 & \textbf{0.742} & 7.52 & 3.39 $\pm$ 0.11 & 2.43 $\pm$ 0.14\\
    \hline
    Baseline               & \multirow{4}{*}{16.2M} & 0.827 & \textbf{1.02} & \textbf{3.82 $\pm$ 0.12} & 3.20 $\pm$ 0.14 & 0.720 & 2.16 & \textbf{3.88 $\pm$ 0.10} & 2.77 $\pm$ 0.14\\
    
    Baseline + CS  & & 0.825 & 1.29 & 3.81 $\pm$ 0.12 & 3.27 $\pm$ 0.15 & 0.682 & 2.75 & 3.55 $\pm$ 0.12 & 2.72 $\pm$ 0.14\\
    
    Baseline + PS          & & 0.814 & 1.10 & 3.72 $\pm$ 0.14 & 3.30 $\pm$ 0.15 & 0.722 & 2.15 & 3.80 $\pm$ 0.10 & 2.72 $\pm$ 0.14\\
    
    Baseline + LF & & \textbf{0.836} & 1.10 & \textbf{3.82 $\pm$ 0.12} & \textbf{3.34 $\pm$ 0.15} & 0.729 & \textbf{2.14} & 3.85 $\pm$ 0.12 & \textbf{2.89 $\pm$ 0.14}\\
    
    \hline
    
    \end{tabular}
    \caption{Objective and subjective zero-shot experiment results of the intra-lingual test that generated English speech samples from English reference speech (En $\rightarrow$ En), and the cross-lingual test that generated English speech samples from Korean reference speech (Ko $\rightarrow$ En). NP indicates a number of parameters. The MOS and SMOS are reported with 95\% CIs. Note that the speech samples of GT and GT-re systems of the cross-lingual test were in Korean.}
    \label{tab:result1}
\end{table*}

\subsection{Results}
The objective and subjective results\footnote{Audio samples can be found online: \\\url{https://srtts.github.io/latent-filling}} are summarized in Table \ref{tab:result1}. The proposed Baseline+LF system exhibited outstanding performance in terms of SECS and SMOS metrics for both intra-lingual (En $\rightarrow$ En) and cross-lingual (Ko $\rightarrow$ En) tests. Specifically, it achieved the highest SMOS and SECS scores in the intra-lingual test, and the best SMOS score along with the second-best SECS score in the cross-lingual test. Compared to the YourTTS system, it achieved 0.55 and 0.46 higher SMOS scores in the intra- and cross-lingual tests, respectively, while employing approximately 60\% fewer parameters.

When compared to the baseline system, our proposed system demonstrated SMOS improvements of 0.14 and 0.12 in the intra-lingual and cross-lingual tests, respectively, as well as SECS improvements of 0.009 for both tests\footnote{The confidence intervals (CIs) for SECS in both the Baseline and Baseline+LF systems were 0.005 and 0.003 for the intra-lingual and cross-lingual tests, respectively.}. These results indicate that by leveraging the LF method to fill the unexplored latent space of the speaker embeddings, the speaker similarity of the ZS-TTS system significantly improved. 
The WER and MOS scores for the intra-lingual and cross-lingual tests were slightly deteriorated, respectively. We suppose that this was because the LF method generated a new speaker embedding, but without the accompanying content information. However, the increase in the WER score for the intra-lingual test was marginally small, and the WER score of the Baseline+LF system remained much lower than that of the GT.

In contrast, the data augmentation approaches (Baseline+CS and Baseline+PS systems) demonstrated mixed results. While the SMOS score improved in the intra-lingual test, all other evaluation metrics deteriorated or slightly improved when compared to the baseline system. This decline was attributed to the low quality of the augmented speech. Conversely, by performing augmentation in the latent space using the LF method, our proposed approach successfully improved speaker similarity without degrading speech quality, in contrast to input-level data augmentation approaches.

\begin{table}[t]
    \small
    \centering
    \begin{tabular}{l|r|r|r}
    \hline
    \textbf{System} & \multicolumn{1}{c|}{\textbf{SECS}}  & \multicolumn{1}{c|}{\textbf{WER} ($\%$)} &
    \multicolumn{1}{c}{\textbf{CSMOS}}\\
    \hline
    Baseline + LF   & 0.836 & \textbf{1.10} & N/A\\
    \hline
    \quad w/o noise adding & \textbf{0.838} & 1.13 & -0.031\\
    \quad w/o interpolation & 0.837 & 1.35 & -0.115 \\
    \hline
    \end{tabular}
    \caption{Objective and CSMOS test results of ablation study on the intra-lingual (En $\rightarrow$ En) test.}
    \label{tab:result2}
\end{table}

\subsection{Ablation study}
To investigate the impact of interpolation and noise addition in the LF method, we constructed separate ZS-TTS systems that utilized LF without noise addition and interpolation. In addition to objective evaluation, we conducted a comparative similarity MOS (CSMOS) test to assess speaker similarity in comparison to the Baseline+LF system. Testers were asked to evaluate the compared systems on a scale ranging from -3 to 3, with an interval of 0.5. The results are presented in Table \ref{tab:result2}.

When each method was omitted, we observed a degradation in both CSMOS and WER scores compared to when both methods were employed. Meanwhile, interpolation proved to be more crucial for performance improvement compared to noise addition. This observation can be attributed to the fact that noise addition, as illustrated in Figure \ref{fig:lf}, typically generates new speaker embeddings that closely resemble existing speaker embeddings. In contrast, interpolation generates new speaker embeddings in larger regions and plays a more vital role in enhancing the system's performance.

\section{Conclusion}
In this paper, we proposed a latent filling (LF) method, which leverages latent space data augmentation to the speaker embedding space of the ZS-TTS system. By introducing the latent filling consistency loss, we successfully integrated LF into the existing ZS-TTS framework seamlessly. Unlike previous data augmentation methods applied to input speech, our LF method improves speaker similarity without compromising the naturalness and intelligibility of the generated speech. 
 
\ninept
\bibliographystyle{IEEEbib}
\bibliography{strings}

\end{document}